\documentclass[11pt,titlepage]{article}
\usepackage[a4paper, dvips=true, scale=0.86, nofoot, centering]{geometry}
\usepackage[T1]{fontenc}

\usepackage{times}

\usepackage{amsmath}
\usepackage{amssymb}

\usepackage{accents}

\usepackage{graphicx}

\usepackage[round,numbers,sort&compress]{natbib}
\title{Decomposition of Complex Reaction Networks into Reactons}

\author{Rapha\"el~Plasson\thanks{
           Corresponding author.  Address:
           Nordita,
       Roslagstullsbacken 23,
       106 91  Stockholm, Sweden,
       Tel.:~+46-8-5537 8718, Fax:~+46-8-5537 8601} \\
    Nordita, \\
    Stockholm, Sweden
    \and Hugues~Bersini \\
    IRIDIA, \\
    ULB, Brussels, Belgium,
    \and Axel~Brandenburg \\
    Nordita, \\
    Stockholm, Sweden}

\date{}

\begin{document}

\maketitle

\abstract{The analysis of complex reaction networks is of great
  importance in several chemical and biochemical fields (interstellar chemistry, prebiotic
  chemistry, reaction mechanism, etc). In this article, we propose
  to simultaneously refine and extend for general chemical reaction systems the
  formalism initially introduced for the description of metabolic networks.
  The classical approaches through the computation
  of the right null space leads to the decomposition of the network
  into complex ``cycles'' of reactions concerned with all metabolites.
  We show how, departing from the left null space computation, the flux analysis can be decoupled
  into linear fluxes and single loops, allowing a more refine qualitative analysis
  as a function of the antagonisms and connections among these local fluxes.
  This analysis is made possible by the decomposition of the molecules into elementary subunits,
  called
"reactons" and the consequent decomposition of the
  whole network into simple first order unary partial reactions related with simple
  transfers of reactons from one molecule to another. This article
  explains and justifies the algorithmic steps leading to the total decomposition of the reaction network
  into its constitutive elementary subpart.

}

\clearpage

\section*{Introduction}

The dynamical analysis of complex reaction networks such as the ones
characterizing interstellar chemistry~\citep{wakelam-06}, complex
reaction mechanism~\citep{ross-99} or metabolic
functions~\citep{papin-04} should be facilitated by algorithmic means
to decouple these networks. It is conceivable that the presence of
interesting dynamical phenomena like bifurcation or symmetry breaking
is mainly due to structural antagonisms between reaction sub-networks
provoking ``threshold'' effects. We propose in this article various
algorithmic tools to allow such decoupling of complex reaction
networks into simpler sub-networks. Any of these sub-networks will be
restrictively concerned with the successive transformations of one
given chemical group that will carry the name of ``reacton'' (in
reminiscence of the organic chemistry ``synthon''), defined as parts
of the molecules that are never broken into smaller pieces by any
reaction of the network (but internal rearrangements of the reactons
are possible).

There has been a great amount of literature dedicated to the analysis
of these reaction networks into sets of balanced reaction fluxes
obtained by computing the right null space (or the ``row'' null space)
of the stoichiometric matrix~\citep{pfeiffer-99}. However such
analysis, while being performed on the complete molecules, leads to
the discovery of fluxes that cross the whole set of reactions, making
difficult the detection of antagonistic subsets and the automatic
anticipation of interesting dynamical phenomena.

The computing of the left null space has been investigated by
\citet{famili-03}. They have shown that this leads to ``pools of
conserved metabolites'' through the reaction network. Similar
descriptions lead to concepts of ``conservation
analysis''~\citep{vallabhajosyula-06} or ``metabolic flux
analysis''~\citep{klamt-03}. Beyond what has been proposed in these
previous studies, this leads to the total decomposition of each
molecule into elementary subunits, the reactons. We argue later in the
article that, such a conservation analysis allows to simplify the global
reaction network into smaller sub-networks.

In order to simplify the reaction network analysis, the first step
proposed in this article is to compute the left null space (or the
``column'' null space). The investigation of an optimal basis of this
null space (this notion of "optimal basis" will be clarified later)
reflects the existence of elementary reactons.  Once the reactons are
identified, the second step is to decouple the complete reaction
network by simply restricting each sub-network to the single reacton
that it is concerned with. Provided the reacton basis is correctly
selected, it leads to substantially smaller networks.

\begin{figure}
   \begin{center}
      \includegraphics*[width=1.5in]{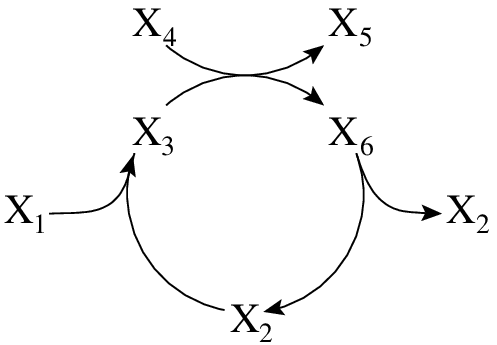}
      \caption{}
      \label{fig:global-loop}
   \end{center}
\end{figure}

Taking for instance the following reaction network:
\begin{eqnarray}
  \label{eq:sys-ex}
  X_1 + X_2 & \xrightarrow{R_1}& X_3 \label{eq:ex1}\\
  X_3 + X_4 & \xrightarrow{R_2} & X_5 + X_6 \label{eq:ex2}\\
  X_6 & \xrightarrow{R_3} & 2 X_2 .\label{eq:ex3}
\end{eqnarray}
the result of the algorithmic analysis performed on this system
should automatically lead to the identification of an
autocatalytic cycle based one the $X_2$, $X_3$ and $X_6$
compounds, as they are being recycled, of two incoming fluxes of
$X_1$ and $X_4$, and two outgoing fluxes of $X_5$ and $X_2$, as
represented in Figure~\ref{fig:global-loop}. It is easy to detect
the presence of two linear fluxes and one cycle, the most
convenient building blocks to perform the following dynamical
analysis (cycles and their intrinsic positive feedback are
responsible for interesting dynamical effects) and to allow the
anticipation of any interesting out-of-equilibrium phenomena such
as the appearance of bifurcations.

\section{Definition of the network}

The studied chemical network is composed of $n$ different compounds
noted $X_i$ for $1 \leq i \leq n$, and $r$ different transformations
noted $R_j$ for $1 \leq j \leq r$. All these reactions are complete
chemical transformations, and are thus mass balanced between reactant
and products. In the example system of Eq.~\ref{eq:sys-ex}, we have
$n=6$ and $r=3$.

Each reaction can be written in the form:
\begin{equation}
  \forall j \in [1,r] \qquad R_j \quad : \quad
  \sum_{i=1}^n \nu_{i,j}X_i=0  .\label{eq:reac_def}
\end{equation}
$\nu_{i,j}$ is the stoichiometric coefficient of the compound $i$ in
the transformation $j$. It is by convention positive for products
(formed compounds), and negative for reactants (disappearing
compounds).

These stoichiometric coefficients can be gathered in a single $n
\times r$ matrix $\overleftrightarrow{\nu}$. In this work, all
matrices will be noted with a
$\overleftrightarrow{\phantom{\nu}}$. The matrix for the system given
in the example is the following $6 \times 3$ matrix:
\begin{equation}
  \label{eq:matr_ex}
  \overleftrightarrow{\nu}=
  \begin{pmatrix}
    -1&0&0\\
    -1&0&+2\\
    +1&-1&0\\
    0&-1&0\\
    0&+1&0\\
    0&+1&-1\\
  \end{pmatrix}.
\end{equation}

This stoichiometric matrix is actually formed by the juxtaposition of
all reaction column vectors:
\begin{equation}
  \overleftrightarrow{\nu} =
  \left(
    \overrightarrow r_1 | \overrightarrow r_2 | \cdots | \overrightarrow r_r
  \right).
\end{equation}
All column vectors will be noted with a $\overrightarrow{\phantom{r}}$,
to be distinguished from the row vectors noted with a
$\overleftarrow{\phantom{c}}$. The purpose of this distinction is to
easily know the dimension of the vectors, $\overrightarrow{a}$ being of
dimension $n\times 1$, and $\overleftarrow{a}$ of dimension $1\times r$.

By noting $\overrightarrow{X}$ as the column vector of the $X_i$,
Eq.~\ref{eq:reac_def} can be written as a matrix multiplication:
\begin{equation}
  \overrightarrow{X}^T\overleftrightarrow{\nu}=\overleftarrow{0},
\end{equation}
or, separating each reaction:
\begin{equation}
 \forall j \in [1,r] \qquad  \overrightarrow{X}^T\overrightarrow{r_j}=0 .
\end{equation}

The traditional analysis of the stoichiometric matrix relies on
the calculation of its null spaces. The right null space, or the
set of combinations of columns of the matrix that gives
$\overrightarrow{0}$  can be expressed as~\citep{meyer-00}:
\begin{eqnarray}
  \overleftarrow{\mathrm{Null}}(\overleftrightarrow{\nu})&=&
  \left\{
    \overleftarrow{b}^T \quad | \quad \overleftrightarrow{\nu}
    \overleftarrow{b}^T
    = \overrightarrow{0}
  \right\}.%
\end{eqnarray}
This corresponds to the set of cycles of reactions, i.e. the
combinations of reactions that results in no global change inside
the system~\citep{pfeiffer-99}.  A base of this null space can be
computed by the Gauss-Jordan elimination~\citep{press-92}, and
represented by:
\begin{equation}
  \overleftarrow{\mathrm{Null}}(\overleftrightarrow{\nu})\quad : \quad
  \overleftrightarrow{\sigma}=
  \left(
    \begin{array}{c}
      \overleftarrow{\sigma_1}\\
      \hline
      \vdots \\
      \hline
      \overleftarrow{\sigma_s}
    \end{array}
  \right)\quad | \quad \overleftrightarrow{\nu}
  \overleftrightarrow{\sigma}^T = \overrightarrow{0}.
\end{equation}
All the $ \overleftarrow{\sigma_k}$ are linearly independent row
vectors of the null space, each one being the row $k$ of the matrix
$\overleftrightarrow{\sigma}$. The total matrix
$\overleftrightarrow{\sigma}$ is thus full rank, and all vectors of
the null space can be expressed as a linear combination of the $
\overleftarrow{\sigma_k}$.

On the other hand, the left null space, i.e. the set of
combinations of rows of the matrix that gives $\overleftarrow{0}$,
can be expressed as~\citep{meyer-00}:
\begin{eqnarray}
  \overrightarrow{\mathrm{Null}}(\overleftrightarrow{\nu})
  &=&
  \overleftarrow{\mathrm{Null}}(\overleftrightarrow{\nu}^T)
  \\
  &=&
  \left\{
    \overrightarrow{a} \quad | \quad \overrightarrow{a}^T \overleftrightarrow{\nu}
    = \overleftarrow{0}
  \right\}.\label{eq:null_molec}
\end{eqnarray}
This left null space of $\overleftrightarrow{\nu}$ is the right
null space of $\overleftrightarrow{\nu}^T$.  It corresponds to the
mass balance in compounds~\citep{famili-03}. Similarly, a base of
this null space can be represented as:
\begin{equation}
  \overrightarrow{\mathrm{Null}}(\overleftrightarrow{\nu})\quad : \quad
  \overleftrightarrow{\sigma}=
\left(
 \overrightarrow{\sigma_1}|\hdots| \overrightarrow{\sigma_s}
\right) \quad | \quad \overleftrightarrow{\sigma}^T   \overleftrightarrow{\nu}
  = \overleftarrow{0},
\end{equation}
where $\overrightarrow{\sigma_k}$ indicates the column $k$ of the
matrix $\overleftrightarrow{\sigma}$. 

\section*{Molecule Decomposition into reactons}

\subsection*{Elementary Reactons}

A reacton is defined as a subpart of a molecule that is never
broken into smaller parts by any of the reactions composing the
network. The chemical reaction network can thus be seen as simple
recombinations of reactons.

A first obvious category of reactons is composed of the atoms, but
larger groups are more likely.  Typically, in a polymerization
system, monomers could be such reactons.

 Molecules can be seen as linear combinations of
reactons such as:
\begin{eqnarray}
  X_i &=& \sum_{k=1}^a \alpha_{i,k}A_k \label{eq:dec_synth}\\
  \Leftrightarrow \quad \overrightarrow{X}&=&\overleftrightarrow{\alpha}\overrightarrow{A},
\end{eqnarray}
with $a$ being the total number of different reactons chosen for
the decomposition. All the $\alpha_{i,k}$ are positive or zero
integers, and represent the number of reacton $A_k$ in $X_i$. The
decomposition of $X_i$ is represented by the vector
$\overrightarrow{\alpha_i}$. The reactons can be represented as a
column vector $\overrightarrow{A}$ of dimension $a$. The matrix
$\overleftrightarrow{\alpha}$ of dimension $n\times a$ represents the decomposition of the whole
set of molecules. 

As molecules can always be at least decomposed into unbreakable atoms
(no nuclear reaction is obviously considered here), there always
exists at least one possible combination of reactons. Nevertheless,
the computation of the left null space should provide us with more
useful reactons in order to simplify the whole network into various
subsets of "independent" reactive pathways.

Any reaction can be written in terms of reactons by substituting
Eq.~\ref{eq:dec_synth} in Eq.~\ref{eq:reac_def}: 
\begin{eqnarray}
  \sum_{i=1}^n\nu_{i,j} \sum_{k=1}^a \alpha_{i,k}A_k   &=& 0 \\
  \sum_{k=1}^a
  \left(
    \sum_{i=1}^n\nu_{i,j}\alpha_{i,k}
  \right)A_k &=& 0\\
  \Leftrightarrow \quad
  \overrightarrow{A}^T\overleftrightarrow{\alpha}^T\overleftrightarrow{\nu} 
  &=&\overleftarrow{0}\label{eq:dec_reac_synth-a}
\end{eqnarray}
There is a mass balance between each reacton $A_k$: they are never
broken into smaller compounds, so that there is always the same number
of every reacton for both reactants and products in any reaction. We
thus have:
\begin{eqnarray}
  \forall j \in [1,r] ~, \forall k \in [1,a] \quad , \quad
  \sum_{i=1}^n\nu_{i,j}\alpha_{i,k}  &=& 0 \label{eq:dec_reac_synth}\\
 \Leftrightarrow \quad \forall k \in [1,a] \quad , \quad \overrightarrow{\alpha_k}^T\overleftrightarrow{\nu}&=&\overleftarrow
  0
\end{eqnarray}
Thus, we have according to  Eq.~\ref{eq:null_molec} and Eq.~\ref{eq:dec_reac_synth}:
\begin{eqnarray}
  \label{eq:nullsp_synth}
 \overrightarrow{\alpha_k} &\in& \overrightarrow{\mathrm{Null}}
  \left(
    \overleftrightarrow{\nu}
  \right)
\end{eqnarray}
All the vectors describing the decomposition of a molecule $X_i$ into
reactons $A_k$, as it can be easily obtained by the atomic
decomposition, are in the left null space of the stoichiometric
matrix.

Let us take the system given in the example of
Eq.~\ref{eq:ex1}-\ref{eq:ex1}, and further decompose it into
small reactons (atoms):
\begin{eqnarray}
  \label{eq:sys-ex-decomp2}
  \mathrm{CH_4O + CH_2O} & \xrightarrow{R_1}&
  \mathrm{C_2H_6O_2} \label{eq:exf1}\\
  \mathrm{C_2H_6O_2 + O_2} & \xrightarrow{R_2} & \mathrm{H_2O_2 + C_2H_4O_2} \\
  \mathrm{C_2H_4O_2} & \xrightarrow{R_3} & 2~ \mathrm{CH_2O}. \label{eq:exf3}
\end{eqnarray}
This system is only for the sake of demonstration and has no real
counterpart. A realistic system will be considered in the penultimate
section.  $X_1=\mathrm{CH_4O}$, $X_2=\mathrm{CH_2O}$,
$X_3=\mathrm{C_2H_6O_2}$, $X_4=\mathrm{O_2}$, $X_5=\mathrm{H_2O_2}$,
and $X_6=\mathrm{C_2H_4O_2}$. There are three atomic reactons:
$A_1=\mathrm{C}$, $A_2=\mathrm{H}$ and $A_3=\mathrm{O}$. The mass
balance in each of these reactons can be seen by the fact that all
reactions are equilibrated. The matrix writing of this is:
\begin{equation}
  \overleftrightarrow{\alpha}=[\overrightarrow{\alpha_1}|\overrightarrow{\alpha_2}|\overrightarrow{\alpha_3}]=
  \begin{pmatrix}
    1&4& 1\\
    1&2& 1\\
    2&6& 2\\
    0&0& 2\\
    0&2& 2\\
    2&4& 2
  \end{pmatrix}
\end{equation}
It can be easily verified that the relation
$\overleftrightarrow{\alpha}^T\overleftrightarrow{\nu}=\overleftarrow{0}$
is satisfied. This simple decomposition of the molecules of the
reaction network into their atoms provides a first possible
solution of the left null space.

\subsection*{Null Space Base and Optimal Reacton Decomposition}

This \emph{a priori} decomposition only gives a collection of reactons, that 
may not describe the whole null space.  It is possible to compute a base of 
the left null space of $\overleftrightarrow{\nu}$ by a Gauss-Jordan
elimination~\citep{press-92}, represented by the matrix
$\overleftrightarrow{\sigma}$:
\begin{equation}
  \overrightarrow{\mathrm{Null}}
  \left(
    \overleftrightarrow{\nu}
  \right)
  \quad : \quad
  \overleftrightarrow{\sigma}
  =
  \left[
    \overrightarrow{\sigma_1} | \ldots  | \overrightarrow{\sigma_s}
  \right]
\end{equation}
all the $\overrightarrow{\sigma_k}$ vectors that are found
respect Eq.~\ref{eq:nullsp_synth} and thus result in one
possible reacton decomposition. There is always a linear mapping
from any set of reactons to the original set found by computing
the null space. It is thus possible to derive any new and more
convenient reacton decomposition from this preliminary set.

Of course, several basis are possible. It is therefore important to
find an ``optimal'' decomposition. As the $\sigma_{i,k}$ represent the
number of reactons present in each molecule, large values of
$\sigma_{i,k}$ imply small reactons (because it means that more
reactons are required to build the molecule). As a consequence,
researching the linear combinations of the $\overrightarrow{\sigma_k}$
that compose the $\overrightarrow{\mathrm{Null}} \left(
  \overleftrightarrow{\nu} \right)$ while minimizing the individual
$\sigma_{i,k}$ and maximizing the number of nought values will lead to
a desired maximization of the size of the reactons. Obtaining a base
of the null space composed of such larger reactons present great
interest, as the larger molecular subparts that are never broken
through the reactions are likely to represent fundamental building
blocks of the system. This should lead to an optimal network
decomposition characterized by a small number of reactons.

Computing the left null space base of the example system, by using
the Gauss pivot method, gives the three following null vectors:
\begin{equation}
  \overrightarrow{\mathrm{Null}}(\overleftrightarrow{\nu}) \quad :
  \quad
    \begin{pmatrix}
      -1&+1&+1\\
      0 & 0&+1\\
      -1&+1&+2\\
      +1&0 & 0\\
      0 &+1& 0\\
      0 & 0&+2
    \end{pmatrix}
\label{eq:first_dec}
\end{equation}
For a good representation of reactons, all the vector elements
must  be  positive or zero integers. A new base can be
searched for, by linearly combining the null vectors, aiming at
maximizing the number of zero elements in each new vector.

The second and third reacton possess only positive values and thus
can be kept as such. In order to eliminate the negative values of
the first one, we can just add the two first reactons:
\begin{equation}
  \overrightarrow{\mathrm{Null}}(\overleftrightarrow{\nu}) \quad :
  \quad \overleftrightarrow{\sigma}
  =[\overrightarrow{\sigma_1}|\overrightarrow{\sigma_2}|\overrightarrow{\sigma_3}]=\begin{pmatrix}
    0 &1&  1\\
    0 &0&  1\\
    0 &1&  2\\
    1 &0&  0\\
    1 &1&  0\\
    0 &0&  2
  \end{pmatrix}
\end{equation}

This decomposition can be seen as a better one than the atomic
decomposition. The molecules are at most decomposed into $3$
reactons, instead of up to $10$. Moreover, two molecules ($X_2$
and $X_4$) are even composed of only one reacton, and can thus be
identified as elementary building blocks of the system.

\subsection*{Elementary decomposition of reactons}

Since we can express any reacton as a linear combination of the
vectors of the null space base $\overleftrightarrow{\sigma}$, it is possible to
link the elementary reactons to the atomic reactons. The
elementary decomposition
$\overleftrightarrow{\alpha}= \left(
    \overrightarrow{\alpha_1} | \ldots  | \overrightarrow{\alpha_a}
  \right)
$, for a system composed of $a$ atoms from $A_1$ to
$A_a$, can be written as:
\begin{equation}
  \overleftrightarrow{\alpha} 
  \begin{pmatrix}
    A_1\\
    \vdots\\
    A_a
  \end{pmatrix}
  =
  \begin{pmatrix}
    X_1\\
    \vdots\\
    X_n
  \end{pmatrix}=\overrightarrow X .\label{eq:atom_decomp}
\end{equation}
The reacton decomposition $\overleftrightarrow{\sigma}$  composed of $s$ reactons
from $S_1$ to
$S_s$, can be written as:
\begin{equation}
  \overleftrightarrow{\sigma} 
  \begin{pmatrix}
    S_1\\
    \vdots\\
    S_s
  \end{pmatrix}
  =\overrightarrow X . \label{eq:reac_decomp}
\end{equation}

Expressing $\overleftrightarrow{\alpha}$ in terms of the null space base is equivalent to finding
a matrix $\overleftrightarrow{T}$ so that:
\begin{equation}
  \overleftrightarrow{\sigma}  \overleftrightarrow{T}=\overleftrightarrow{\alpha}
\end{equation}
Combining Eq.~\ref{eq:atom_decomp} and Eq.~\ref{eq:reac_decomp} gives:
\begin{eqnarray}
  \overleftrightarrow{\alpha} \overleftrightarrow{A}
  &=&  \overleftrightarrow{\sigma} \overleftrightarrow{S}\\
  \overleftrightarrow{\sigma}  \overleftrightarrow{T} \overleftrightarrow{A}
  &=&  \overleftrightarrow{\sigma} \overleftrightarrow{S}.
\end{eqnarray}
Because $\overleftrightarrow{\sigma}$ is full rank, it is possible to obtain the
decomposition of the reactons into atom by:
\begin{equation}
\overleftrightarrow{S}=   \overleftrightarrow{T} \overleftrightarrow{A}.
\end{equation}

In the example of Eq.~\ref{eq:exf1}-\ref{eq:exf3}, linear combinations
between $\overrightarrow{\alpha_k}$ and $\overrightarrow{\sigma_k}$
can easily be found:
\begin{eqnarray}
    \overrightarrow{\alpha_1}&=&\overrightarrow{\sigma_3} \\
    \overrightarrow{\alpha_2}&=&2\overrightarrow{\sigma_2}+2\overrightarrow{\sigma_3}\\
    \overrightarrow{\alpha_3}&=&2\overrightarrow{\sigma_1}+\overrightarrow{\sigma_3},
\end{eqnarray}
leading to:
\begin{eqnarray}
  \overleftrightarrow{\sigma} 
  \begin{pmatrix}
    0&0&2\\
    0&2&0\\
    1&2&1
  \end{pmatrix}
  &=&\overleftrightarrow{\alpha}\\
  \begin{pmatrix}
    S_1\\
    S_2\\
    S_3
  \end{pmatrix}
  &=&
  \begin{pmatrix}
    0&0&2\\
    0&2&0\\
    1&2&1
  \end{pmatrix}
  \begin{pmatrix}
    \mathrm{C}\\
    \mathrm{H}\\
    \mathrm{O}
  \end{pmatrix}.
\end{eqnarray}

As a consequence, this new decomposition, aiming for bigger reactons,
gives the three following ones: $S_1=\mathrm{O_2}$, $S_2=\mathrm{H_2}$
and $S_3=\mathrm{CH_2O}$. We thus have $X_1=\mathrm{(H_2)(CH_2O)}$,
$X_2=\mathrm{(CH_2O)}$, $X_3=\mathrm{(H_2)(CH_2O)_2}$,
$X_4=\mathrm{(O_2)}$, $X_5=\mathrm{(O_2)(H_2)}$,
$X_6=\mathrm{(CH_2O)_2}$. It can be checked that all the reactions of
the example can be written as association and dissociation of these
only three sub-elements.

\section*{Transformation decomposition}

It is thus possible to find an automatic decomposition of molecules into
reactons, and to find the atomic decomposition of these reactons into
atoms, as long as the formula of molecules is known.

The decomposition of molecules $X_i$ into $s$ reactons $S_k$ can then be applied to
the transformations themselves, allowing a focused study of each
kind of reacton inside the system and how they do interfere. The
molecules are decomposed as:
\begin{equation}
  X_i = \sum_{k=1}^s \sigma_{i,k}X_i^{(k)} \label{eq:dec_react}.
\end{equation}
At this point, it is important to track the position of reactons
in the molecules they belong to. $X_i^{(k)}$ represents the
subpart of a molecule $X_i$ that corresponds to a reacton $S_k$.

We can decompose each reaction $R_j$ into $s$ partial reactions, each
one describing the transfer of reactons from one compound to another:
\begin{eqnarray}
    R_j & : & \sum_{k=1}^s
    \sum_{i=1}^n\nu_{i,j}\sigma_{i,k}X_i^{(k)}   = 0\\
     R_j & : &  \sum_{k=1}^s R_j^{(k)}    = 0.
\end{eqnarray}
Because there is a mass balance in each reacton $S_k$, we can
decompose the reactions into partial reactions relative to each reacton:
\begin{eqnarray}
    R_j^{(k)} & : & \sum_{i=1}^n\nu_{i,j}\sigma_{i,k}X_i^{(k)}   = 0\\
    \overrightarrow{r_j}^{(k)} &=& \overrightarrow{\sigma_k} \circ \overrightarrow{r_j}.
\end{eqnarray}
$R_j^{(k)}$ represents the subpart of the reaction $R_j$ that involves only
the reactons $S_k$. The symbol $\circ$ represent the
Hadamard product (i.e. elementwise product).

The transformations relative to the reacton $S_k$ can be summarized by
the following stoichiometric matrix:
\begin{eqnarray}
  \overleftrightarrow{\nu}^{(k)} &=&
  \left[
    \nu_{i,j}\sigma_{i,k}
  \right]_{i\in[1,n],j\in[1,r]}\\
 &=& (\overrightarrow{\sigma_k}\overleftarrow 1)\circ\overleftrightarrow{\nu}.
\end{eqnarray}
This operation correspond to multiplying each row by the corresponding
element of the reacton vector. As each reacton vector is in the left
null space of the stoichiometric matrix, we have:
\begin{eqnarray}
  \sum_{i=1}^n\nu_{i,j}\sigma_{i,k}&=&0\\   
\Rightarrow \quad \overrightarrow{1}^T\overleftrightarrow{\nu}^{(k)}&=&\overleftarrow{0}\label{eq:cons_react}.
\end{eqnarray}
That is the sum of the rows of each $\overleftrightarrow{\nu}^{(k)}$
is null or, said differently, that the vector $\overrightarrow{1}$ is
in the left null space of each one of these matrices.

The system given in example can be decomposed into the following
three reacton stoichiometric matrices:
\begin{equation}
  \overleftrightarrow{\nu}^{(1)} =
  \begin{pmatrix}
    0&0&0\\
    0&0&0\\
    0&0&0\\
    0&-1&0\\
    0&+1&0\\
    0&0&0
  \end{pmatrix}
  \quad ; \quad
  \overleftrightarrow{\nu}^{(2)} =
  \begin{pmatrix}
    -1&0&0   \\
    0&0&0    \\
    +1&-1&0    \\
    0&0&0    \\
    0&+1&0     \\
    0&0&0
  \end{pmatrix}
  \quad ; \quad
  \overleftrightarrow{\nu}^{(3)} =
  \begin{pmatrix}
    -1&0&0\\
    -1&0&+2\\
    +2&-2&0\\
    0&0&0\\
    0&0&0\\
    0&+2&-2
  \end{pmatrix}.
\end{equation}

A more compact notation can be used. Submatrices can be extracted
for each reacton by removing the reactions that do not involve
this reacton (i.e. $\overrightarrow 0$ columns) and the molecules that do not
contain it (i.e. $\overleftarrow 0$ lines):
\begin{equation}
  \overleftrightarrow{\nu}'^{(1)} =
  \bordermatrix{
    &R_2^{(1)}\cr
    X_4^{(1)}&-1\cr
    X_5^{(1)}&+1\cr
  }
  \quad ; \quad
  \overleftrightarrow{\nu}'^{(2)} =
  \bordermatrix{
    &R_1^{(2)}&R_{2}^{(2)}\cr
    X_{1}^{(2)}&-1&0\cr
    X_{3}^{(2)}&+1&-1\cr
    X_{5}^{(2)}&0&+1
  }
  \quad ; \quad
  \overleftrightarrow{\nu}'^{(3)} =
  \bordermatrix{
    &R_{1}^{(3)}&R_{2}^{(3)}&R_{3}^{(3)} \cr
    X_{1}^{(3)}&-1&0&0\cr
    X_{2}^{(3)}&-1&0&+2\cr
    X_{3}^{(3)}&+2&-2&0\cr
    X_{6}^{(3)}&0&+2&-2
  }.
\end{equation}
Given this simplified notation, we need to keep track of the
meaning of each line and column for not losing the information
about the involved reactions and compounds, but it definitely
gives a simpler view of the reacton subsystem.

\section*{Decomposition into fluxes and cycles}

Reactons are never broken into pieces, they are just transferred
from one molecule to another. From this point on, all the
reactions can be decomposed into simple $A \rightarrow B$
transformation reactions. The whole system can be decomposed into
either linear fluxes (a succession of transformation $A
\rightarrow B \rightarrow C \rightarrow \cdots$) or cyclic fluxes
(a succession of transformations $A \rightarrow B \rightarrow
\cdots \rightarrow A$). This property of the new subsystems will
greatly ease their analysis.

\subsection*{Order of the reactions}

Because of the conservation of the reactons, the sum of the components
of each colon is null (Eq.~\ref{eq:cons_react}). The sum of the
positive numbers is thus identical in absolute value to the sum of
negative numbers. This absolute value gives the number of reactons
engaged in the reaction that is the order of the reaction for a given
reacton. In order to linearize the system, it is necessary to
determine the order of each reaction, so that each reaction of order
$n$ can be divided into $n$ partial reactions of order $1$.

By defining the operations $()^+$ and $()^-$ as follows:
\begin{equation}
  (x)^+ =
  \begin{cases}
    x & \mathrm{if\ } x>0\\
    0 & \mathrm{if\ } x \leq 0
  \end{cases} \qquad ; \qquad
  (x)^- =
  \begin{cases}
    0 & \mathrm{if\ } x \geq 0\\
    -x & \mathrm{if\ } x < 0
  \end{cases}
\end{equation}
We can define the following function, giving the order of each
reaction:
\begin{eqnarray}
  \overleftarrow{\mathrm{ord}}_+\left(\overleftrightarrow{\nu}^{\left(k\right)}\right)
  &=&\left[\sum_{i=1}^n
    \left(\nu_{i,j}\sigma_{i,k}\right)^+\right]_{1 \leq j \leq r}\\
  \overleftarrow{\mathrm{ord}}_-\left(\overleftrightarrow{\nu}^{\left(k\right)}\right)&=&\left[\sum_{i=1}^n
    \left(\nu_{i,j}\sigma_{i,k}\right)^-\right]_{1 \leq j \leq r}\\
  \overleftarrow{\mathrm{ord}}\left(\overleftrightarrow{\nu}^{\left(k\right)}\right)&=&
  \overleftarrow{\mathrm{ord}}_+\left(\overleftrightarrow{\nu}^{\left(k\right)}\right) =   \overleftarrow{\mathrm{ord}}_-\left(\overleftrightarrow{\nu}^{\left(k\right)}\right)
\end{eqnarray}
In our example, we obtain:
\begin{equation}
  \overleftarrow{\mathrm{ord}}\left(\overleftrightarrow{\nu}^{(1)}\right)=
  \begin{pmatrix}
    0&1&0
  \end{pmatrix}
    \quad ; \quad
  \overleftarrow{\mathrm{ord}}\left(\overleftrightarrow{\nu}^{(2)}\right)=
  \begin{pmatrix}
    1&1&0
  \end{pmatrix}
  \quad ; \quad
  \overleftarrow{\mathrm{ord}}\left(\overleftrightarrow{\nu}^{(3)}\right)=
  \begin{pmatrix}
    2&2&2
  \end{pmatrix}
\end{equation}

\subsection*{Connection of the system to external fluxes}

Each combination $\overleftarrow c$ of reactions leads to a
global transformation $\overrightarrow t= \overleftrightarrow{\nu}
 \overleftarrow{c}^T$. Provided this resultant
transformation $\overrightarrow t$ can be compensated by an
external flux  $\overrightarrow f = -\overrightarrow t$, this
combination can be maintained in a steady state. The
stoichiometric matrix can then be extended by multiplying each
reaction according to $\overleftarrow c$, and adding  a new
column representing this flux:
\begin{eqnarray}
  \accentset{\circ}{\overleftrightarrow{\nu}} &=&
   \left[
     (\overrightarrow 1  \overleftarrow c)\circ\overleftrightarrow{\nu} | -\overleftrightarrow{\nu} 
     \overleftarrow{c}^T
   \right]\\
  \accentset{\circ}{\overleftrightarrow{\nu}}^{(k)} &=&
   \left[
     (\overrightarrow 1  \overleftarrow c)\circ\overleftrightarrow{\nu}^{(k)} | -\overleftrightarrow{\nu}^{(k)} 
     \overleftarrow{c}^T
   \right]
\end{eqnarray}
We now obtain matrices whose sum of components of each row
($\accentset{\circ}{\overleftrightarrow{\nu}}\overleftarrow{1}^T$
and
$\accentset{\circ}{\overleftrightarrow{\nu}}^{(k)}\overleftarrow{1}^T$)
is null.

Provided this addition of some external fluxes of reactons, the
resulting system can be maintained in an active steady state. It
is then possible to compute:
\begin{eqnarray}
   \overrightarrow{\mathrm{ord}}(\accentset{\circ}{\overleftrightarrow{\nu}}^{(k)})
  &=&
  \overrightarrow{\mathrm{ord}}_+(\accentset{\circ}{\overleftrightarrow{\nu}}^{(k)})
  =
  \overrightarrow{\mathrm{ord}}_-(\accentset{\circ}{\overleftrightarrow{\nu}}^{(k)})
\end{eqnarray}
Each reacton turns out to be involved in a same number of
creations and destructions.

In the system given as example, we will allow exchanges of compounds
$X_1$, $X_2$, $X_4$ and $X_5$, while compounds $X_3$ and $X_6$ will be
considered as internal compounds.  $\overleftrightarrow{\nu}$ is
written as in Eq.~\ref{eq:matr_ex}.  We first need to identify the
vector $\overleftarrow c$ in order to guarantee the conservation of
internal compounds. We also need to identify the vector
$\overrightarrow t$ in order to guarantee the conservation of the
external ones. Since $X_3$ is only present in reactions $R_1$ and
$R_2$, its conservation demands to combine a same number of $R_1$ and
$R_2$. $X_6$ is only present in reactions $R_2$ and $R_3$, so that a
same number of $R_2$ and $R_3$ is required. As a consequence, there is
only one possible combination of reactions that can lead to a steady
state, $\overleftarrow c=
\begin{pmatrix}
  1&1&1
\end{pmatrix}
$. We then obtain:
\begin{equation}
  \accentset{\circ}{\overleftrightarrow{\nu}}=
  \begin{pmatrix}
    -1&0&0&+1\\
    -1&0&+2&-1\\
    +1&-1&0&0\\
    0&-1&0&+1\\
    0&+1&0&-1\\
    0&+1&-1&0
  \end{pmatrix} \qquad \mathrm{,} \qquad
  \overrightarrow{\mathrm{ord}}
  \left(
    \accentset{\circ}{\overleftrightarrow{\nu}}
  \right)
  =
  \begin{pmatrix}
    1\\
    2\\
    1\\
    1\\
    1\\
    1
  \end{pmatrix} \label{eq:ordsys}
\end{equation}
And for each reacton:

\begin{eqnarray}
  \accentset{\circ}{\overleftrightarrow{\nu}}^{\left(1\right)}&=&
  \begin{pmatrix}
    0&0&0&0\\
    0&0&0&0\\
    0&0&0&0\\
    0&-1&0&+1\\
    0&+1&0&-1\\
    0&0&0&0
  \end{pmatrix} \qquad \mathrm{,} \qquad
  \overrightarrow{\mathrm{ord}}
  \left(
    \accentset{\circ}{\overleftrightarrow{\nu}}^{\left(1\right)}
  \right)
  =
  \begin{pmatrix}
    0\\
    0\\
    0\\
    1\\
    1\\
    0
  \end{pmatrix}\\
  \overleftarrow{\mathrm{ord}}\left(\accentset{\circ}{\overleftrightarrow{\nu}}^{\left(1\right)}\right)&=&
  \begin{pmatrix}
    0& \phantom{+}1& 0& \phantom{+}1
  \end{pmatrix}
\end{eqnarray}
 Denoting by $F$ the part of the matrix responsible for the external fluxes,
 the reduced expression of the matrix is:
 \begin{equation}
   \accentset{\circ}{\overleftrightarrow{\nu}}'^{\left(1\right)}=
   \bordermatrix{
     &R_{2}^{(1)}&F^{\left(1\right)}\cr
     X_{4}^{(1)}&-1&+1\cr
     X_{5}^{(1)}&+1&-1\cr
   }
 \end{equation}

\begin{eqnarray}
  \accentset{\circ}{\overleftrightarrow{\nu}}^{(2)}&=&
  \begin{pmatrix}
    -1&0&0&+1\\
    0&0&0&0\\
    +1&-1&0&0\\
    0&0&0&0\\
    0&+1&0&-1\\
    0&0&0&0
  \end{pmatrix} \qquad \mathrm{,} \qquad
  \overrightarrow{\mathrm{ord}}
  \left(
    \accentset{\circ}{\overleftrightarrow{\nu}}^{(2)}
  \right)
  =
  \begin{pmatrix}
    1\\
    0\\
    1\\
    0\\
    1\\
    0
  \end{pmatrix}\\
  \overleftarrow{\mathrm{ord}}\left(\accentset{\circ}{\overleftrightarrow{\nu}}^{(2)}\right)&=&
  \begin{pmatrix}
     \phantom{+}1& \phantom{+}1& 0& \phantom{+}1
  \end{pmatrix}
\end{eqnarray}
Similarly, the reduced expression of this second reacton matrix
is:

\begin{equation}
  \accentset{\circ}{\overleftrightarrow{\nu}}'^{(2)} =
  \bordermatrix{
    &R_{1}^{(2)}&R_{2}^{(2)}&F^{(2)}\cr
    X_{1}^{(2)}&-1&0&+1\cr
    X_{3}^{(2)}&+1&-1&0\cr
    X_{5}^{(2)}&0&+1&-1
  }
\end{equation}

\begin{eqnarray}
  \accentset{\circ}{\overleftrightarrow{\nu}}^{(3)}&=&
  \begin{pmatrix}
    -1&0&0&+1\\
    -1&0&+2&-1\\
    +2&-2&0&0\\
    0&0&0&0\\
    0&0&0&0\\
    0&+2&-2&0
  \end{pmatrix} \qquad \mathrm{,} \qquad
  \overrightarrow{\mathrm{ord}}
  \left(
    \accentset{\circ}{\overleftrightarrow{\nu}}^{(3)}
  \right)
  =
  \begin{pmatrix}
    1\\
    2\\
    2\\
    0\\
    0\\
    2
  \end{pmatrix}\\
  \overleftarrow{\mathrm{ord}}\left(\accentset{\circ}{\overleftrightarrow{\nu}}^{(3)}\right)&=&
  \begin{pmatrix}
     \phantom{+}2& \phantom{+}2& \phantom{+}2& \phantom{+}1
  \end{pmatrix}
\end{eqnarray}
 The reduced expression of the third reacton matrix is:
\begin{equation}
  \accentset{\circ}{\overleftrightarrow{\nu}}'^{(3)} =
  \bordermatrix{
    &R_{1}^{(3)}&R_{2}^{(3)}&R_{3}^{(3)}&F^{(3)} \cr
    X_{1}^{(3)}&-1&0&0&+1\cr
    X_{2}^{(3)}&-1&0&+2&-1\cr
    X_{3}^{(3)}&+2&-2&0&0\cr
    X_{6}^{(3)}&0&+2&-2&0
  }
\end{equation}

\subsection*{Cycle decomposition}

It is always possible to decompose the system into unitary
transformations. All molecules can be decomposed into reactons and
every reaction can be divided into partial sub-reactions concerned
with the transfer of one reacton from one molecule to another.
This amounts to an extension of the reacton matrix by duplicating
each line and row according to their respective order.

\begin{figure}
   \begin{center}
      \includegraphics*[width=6.5in]{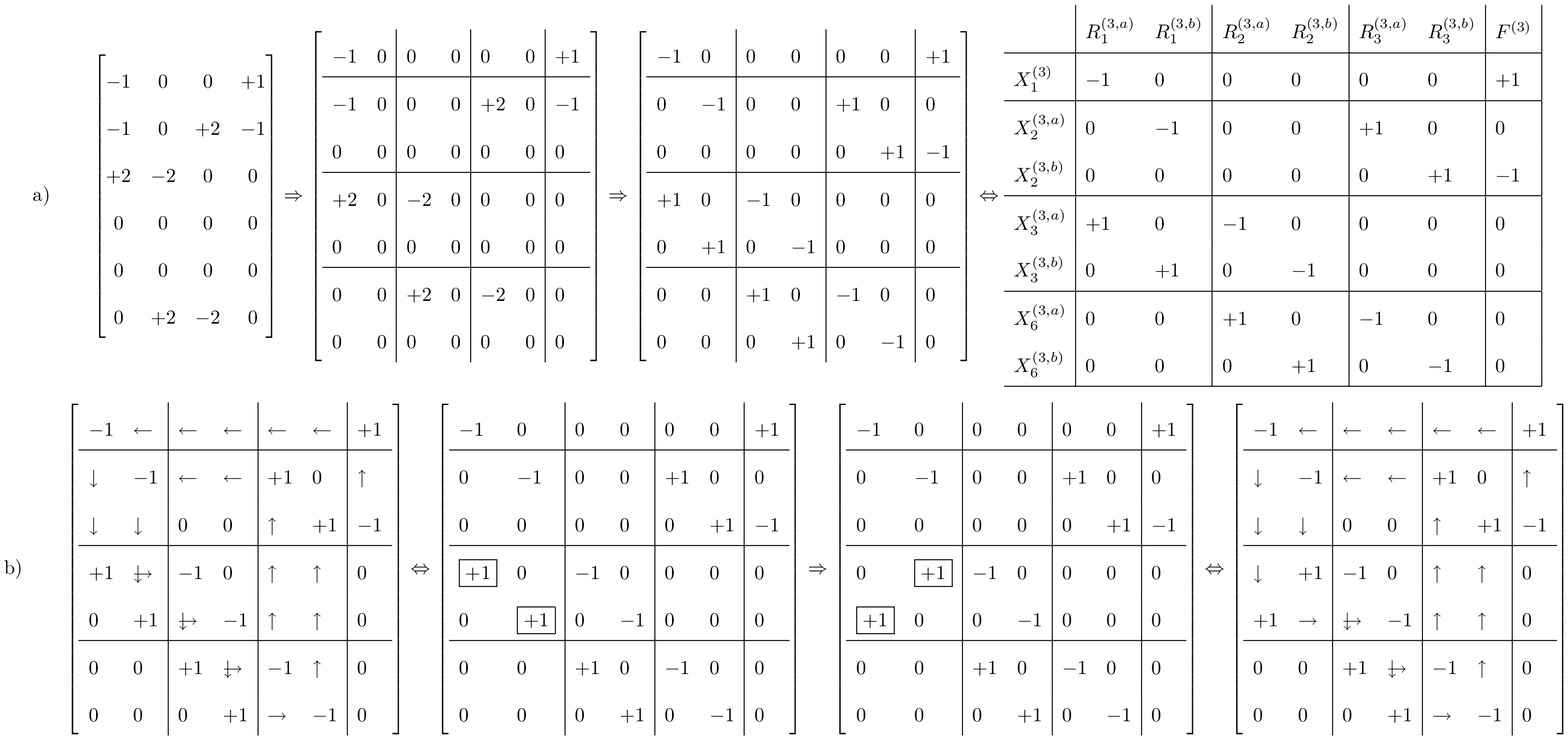}
      \caption{}
      \label{fig:mat-decomp}
   \end{center}
\end{figure}

For instance, the decomposition of
$\accentset{\circ}{\overleftrightarrow{\nu}}^{(3)}$ (i.e. the
matrix associated with the third reacton) is detailed in
Fig.~\ref{fig:mat-decomp}.a and goes as follows. The elements of
the matrix are distributed in such a way that only unitary
reactions are obtained and every reacton is involved in only one
reaction as a reactant and only one reaction as a product. The
first step consists in supplying additional null rows and column
according to the orders of molecules and reactions of the
concerned reacton. In the second step, the matrix is parsed, first
from left to right, then from the top to the bottom. If a non zero
number is found in a column that already contains another number
of the same sign, it is moved on the right. Then, if a non zero
number is found in a row that already contains another number of
the same sign, it is moved down. A number greater than $1$ or
smaller than $-1$ must be considered as several $1$ or $-1$
occupying a same matrix cell. Consequently, this cell is adapted
so as to contain only $1$ and $-1$ located as described above (no
number of the same sign in a same row or column).

A new matrix is thus obtained in which every line and every column
contains one and only one pair of $-1$ and $+1$. As a result, the
molecules have been decomposed into single reactons and the whole
reaction network into single reacton transformations from one
molecule to another. In this new matrix, the $X_{i}^{(k,x)}$ are
associated with one of the reactons $S_k$ of the molecule $X_i$.
Since these apparently different reactons turn out to be
identical, they can be swapped without altering the system.
$F^{(k)}$ is the flux of reacton $S_k$, and $R_{j}^{(k,x)}$ are
the partial reaction $x$ of the reaction $R_{j}^{(k)}$,
corresponding to one single conversion of a reacton $S_k$ from one
molecule to another through the reaction $R_j$. Various
combinations are actually possible, and the different
$R_{j}^{(k,x)}$ can exchange their reactants or their products
without altering the system.

The right null space of this new matrix can simply be computed as
the addition of single columns (Fig.~\ref{fig:mat-decomp}.b). For
a given reacton matrix, there is a simple way to proceed.
Starting from a non zero value, one can go to the other non-zero
value of the same line, then to the same non-zero value of the
same column, then to the same non-zero value of the same line,
etc. When the starting value is reached, the sum of all the
visited columns is $\overrightarrow 0$, and the sum of all the
visited rows is $\overleftarrow 0$. The following part of the
right null space gives rise to a cycle: {\small
\begin{equation}
  X_{1}^{(3)}  \xrightarrow{R_{1}^{(3,a)}} X_{3}^{(3,a)}
  \xrightarrow{R_{2}^{(3,a)}} X_{6}^{(3,a)}  \xrightarrow{R_{3}^{(3,a)}}
  X_{2}^{(3,a)}  \xrightarrow{R_{1}^{(3,b)}} X_{3}^{(3,b)}
  \xrightarrow{R_{2}^{(3,b)}} X_{6}^{(3,b)}  \xrightarrow{R_{3}^{(3,b)}}
  X_{2}^{(3,b)}  \xrightarrow{F^{(3)}} (X_{1}^{(3)})
\end{equation}}
This one unique cycle involves all reactons and reactions. We can
notice that, for example, the cycle goes twice through the
molecule $X_3$, to perform the reaction $R_2$. Since the two
reactons are actually identical inside this molecule, they can be
exchanged. This amounts to the substitution:
\begin{eqnarray}
  \cdots\xrightarrow{R_{1}^{(3,a)}} X_{3}^{(3,a)}
  &\xrightarrow{R_{2}^{(3,a)}}& X_{6}^{(3,a)}\cdots\\
  \cdots\xrightarrow{R_{1}^{(3,b)}} X_{3}^{(3,b)}
  &\xrightarrow{R_{2}^{(3,b)}}& X_{6}^{(3,b)}\cdots
\end{eqnarray}
with:
\begin{eqnarray}
  \cdots\xrightarrow{R_{1}^{(3,a)}} X_{3}^{(3,a)}
  &\xrightarrow{R_{2}^{(3,b)}}& X_{6}^{(3,b)} \cdots \\
  \cdots \xrightarrow{R_{1}^{(3,b)}} X_{3}^{(3,b)}
  &\xrightarrow{R_{2}^{(3,a)}}& X_{6}^{(3,a)} \cdots
\end{eqnarray}
The new version of the decomposition leads to two different
combinations of reactions in the null space:
\begin{eqnarray}
\left[
   \begin{array}{ll|ll|ll|l}
     1&0&0&1&0&1&1
   \end{array}
 \right]&=&R_1^{(3,a)}+R_2^{(3,b)}+R_3^{(3,a)}+F^{(3)} \\
\left[
   \begin{array}{ll|ll|ll|l}
     0&1&1&0&1&0&0
   \end{array}
 \right]&=&R_1^{(3,b)}+R_2^{(3,a)}+R_3^{(3,b)}
\end{eqnarray}
They are complementary (their sum is $\overleftarrow c$) and
there is never more than one partial reaction involved. The system
is thus completely decoupled, keeping apart all particular
reactive fluxes. If there were still crossings like the one just
resolved, a new matrix should be formed by similarly inverting the
crossing point. This process can be continued as long as crossings
still exist, until a fully decomposed reaction network is reached.
With such a decomposition procedure, the following cycles are
obtained:
\begin{equation}
  X_{1}^{(3)}  \xrightarrow{R_{1}^{(3,a)}} X_{3}^{(3,b)}
  \xrightarrow{R_{2}^{(3,b)}} X_{6}^{(3,b)}  \xrightarrow{R_{3}^{(3,b)}}
  X_{2}^{(3,b)}  \xrightarrow{F^{(3)}} (X_{1}^{(3)}) \label{eq:first_cyc}
\end{equation}
\begin{equation}
  X_{2}^{(3,a)}  \xrightarrow{R_{1}^{(3,b)}} X_{3}^{(3,a)}
  \xrightarrow{R_{2}^{(3,a)}} X_{6}^{(3,a)}  \xrightarrow{R_{3}^{(3,a)}}
   (X_{2}^{(3,a)})
\end{equation}
Moreover, $F^{(3)}$ actually represents an incoming flux of $X_{1}^{(3)}$
(noted $F^{(3)}_+$) and an outgoing flux of $X_{2}^{(3,b)}$ (noted
$F_-{(3)}$), so that the cycle of Eq.~\ref{eq:first_cyc} represents a linear
flux, which can be written:
\begin{equation}
  \xrightarrow{F_+^{(3)}}
  X_{1}^{(3)}  \xrightarrow{R_{1}^{(3,a)}} X_{3}^{(3,b)}
  \xrightarrow{R_{2}^{(3,b)}} X_{6}^{(3,b)}  \xrightarrow{R_{3}^{(3,b)}}
  X_{2}^{(3,b)} \xrightarrow{F_-^{(3)}}
\end{equation}

\begin{figure}
   \begin{center}
      \includegraphics*[width=3.25in]{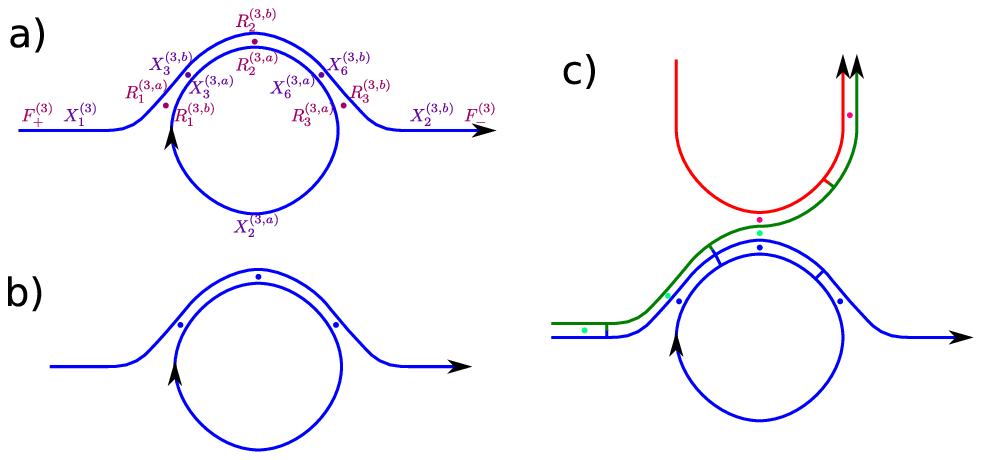}
      \caption{}
      \label{fig:loops}
   \end{center}
\end{figure}

These two fluxes are coupled since the involved partial reactions
are always working together (e.g. $R_{1}^{(3)}=R_{1}^{(3,a)}+R_{1}^{(3,b)}$),
and the two pairs of reactons $\{X_{3}^{(3,a)},X_{3}^{(3,b)}\}$ and
$\{X_{6}^{(3,a)},X_{6}^{(3,b)}\}$ are linked into the same molecules $X_3$
and $X_6$. As $X_2$ is only composed of one $S_3$ reacton,
$X_{2}^{(3,a)}$ and $X_{2}^{(3,b)}$ are not linked, but represent two
reactons that are present in two different $X_2$ molecules. This
can be represented in a graphic form (Fig.~\ref{fig:loops}a)
or, more simply, by just emphasising the fluxes (Fig.~\ref{fig:loops}b).
The linear flux and the cycle are coupled by the three reactions.

The treatment of the two other reactons is much simpler. In each
case, only one linear flux is observed.

For the first reacton:
\begin{equation}
   \xrightarrow{F_+^{(1)}} X_{4}^{(1)}  \xrightarrow{R_{2}^{(1)}} X_{5}^{(1)}
 \xrightarrow{F_-^{(1)}}
\end{equation}

For the second reacton:
\begin{equation}
   \xrightarrow{F_+^{(2)}} X_{1}^{(2)}  \xrightarrow{R_{1}^{(2)}} X_{3}^{(2)}
  \xrightarrow{R_{2}^{(2)}} X_{5}^{(2)} \xrightarrow{F_-^{(2)}}
\end{equation}

\section*{Rebuilding the complete system}

According to $\overrightarrow{\mathrm{ord}} \left(
  \accentset{\circ}{\overleftrightarrow{\nu}} \right)$ (Eq.~\ref{eq:ordsys}), the system is composed of $1 X_1+2 X_2+1 X_3+1 X_4+ 1 X_5+1 X_6$. These
molecules are decomposed as follows into the three reactons:
\begin{eqnarray}
  X_1&=&X_{1}^{(2)}+X_{1}^{(3)}\\
  X_{2}^{(a)}&=&X_{2}^{(3,a)}\\
  X_{2}^{(b)}&=&X_{2}^{(3,b)}\\
  X_3&=&X_{3}^{(2)}+X_{3}^{(3,a)}+X_{3}^{(3,b)}\\
  X_4&=&X_{4}^{(1)}\\
  X_5&=&X_{5}^{(2)}+X_{5}^{(1)}\\
  X_6&=&X_{6}^{(3,a)}+X_{6}^{(3,b)}
\end{eqnarray}
With respect to the flux analysis, the system is represented by
$\overleftarrow c=
\begin{pmatrix}
  1&1&1
\end{pmatrix}
$, that is $1 R_1+1 R_2+1 R_3$. These
reactions are decomposed into the following partial reactions:
\begin{eqnarray}
  R_1&=&R_{1}^{(2)}+R_{1}^{(3,a)}+R_{1}^{(3,b)}\\
  R_2&=&R_{2}^{(1)}+R_{2}^{(2)}+R_{2}^{(3,a)}+R_{2}^{(3,b)}\\
  R_3&=&R_{3}^{(3)}
\end{eqnarray}

The whole system can then be rebuilt by associating the
corresponding reactons and the partial reactions, leading to  Fig.~\ref{fig:loops}c.
The flux of $S_1$ is represented in red, the flux of $S_2$ in
green, and both the flux and cycle of $S_3$ in blue. The dots
indicate the reactions where the fluxes are coupled. The link
between the fluxes represent the molecules, composed of several
reactons.

The total system has thus been decomposed in a way that emphasizes
the evolution of its different subparts. There is a global flux
from $X_1$ to $X_2$, another one from $X_4$ to $X_5$, and an
internal cycle of reacton $S_3$. The autocatalytic property of
this system can be seen by the coupling of a linear flux and an
internal cycle concerning the same reacton $S_3$.

\section*{Application to a Realistic Network}

\begin{figure}
   \begin{center}
      \includegraphics*[width=3.25in]{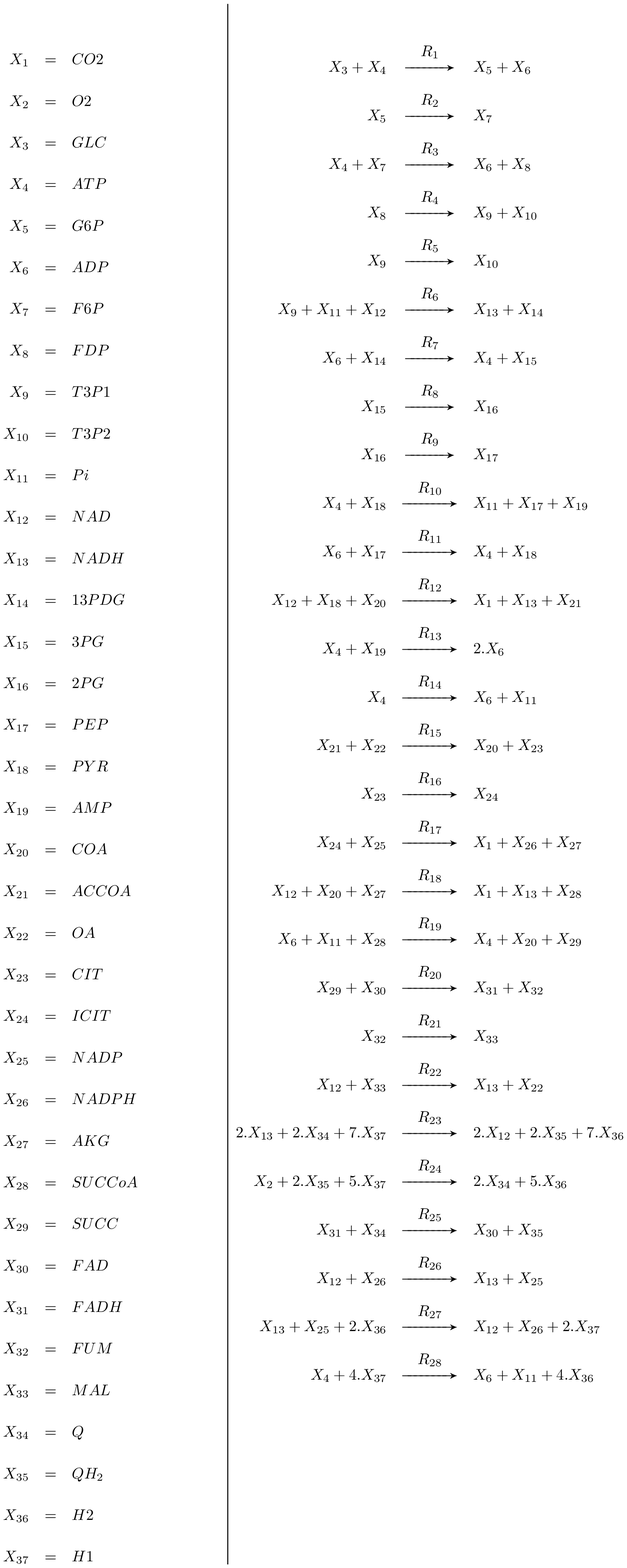}
      \caption{}
      \label{fig:ecoli-metab}
   \end{center}
\end{figure}

\begin{figure}
   \begin{center}
      \includegraphics*[width=6.5in]{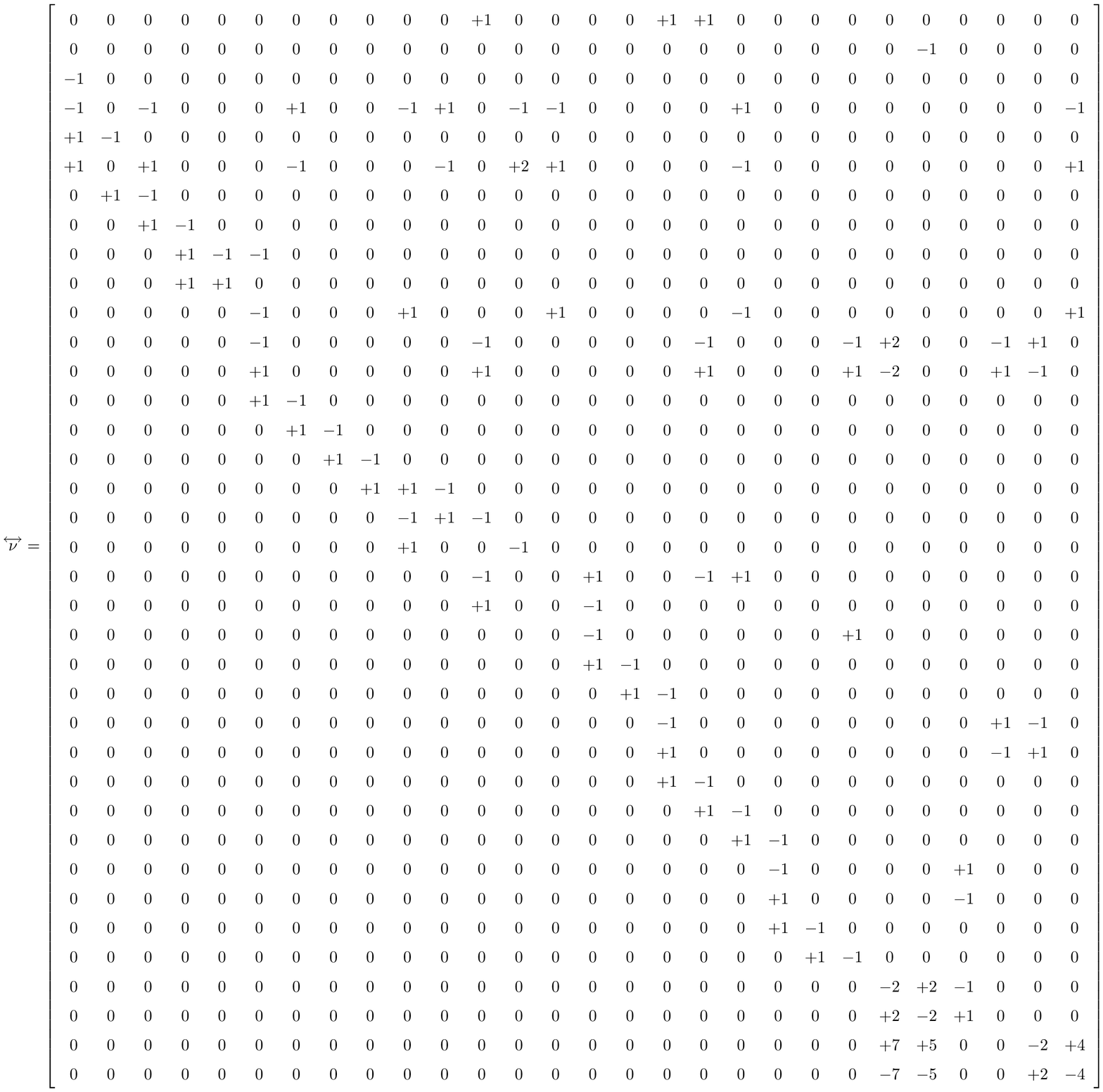}
      \caption{}
      \label{fig:ecoli-stoich}
   \end{center}
\end{figure}

In this section, a more complex and realistic example, a partial
E.Coli metabolism~\citep{beard-04} is treated by applying the same
sequence of algorithmic operations. This system describes the
decomposition of glucose into carbon dioxide. It involves the
transformation of $37$ molecules through $28$ reactions. We have
kept the same notations as described by~\citet{beard-04}, except
for $HEXT$ (exchange of $H^+$ through a membrane) that has been
replaced by two compounds $H1$ and $H2$, corresponding to internal
and external protons, in order to keep the mass balance. This
system is described in Fig.~\ref{fig:ecoli-metab}, and the
corresponding stoichiometric matrix in
Fig.~\ref{fig:ecoli-stoich}.

\begin{figure}
   \begin{center}
      \includegraphics*[width=3.25in]{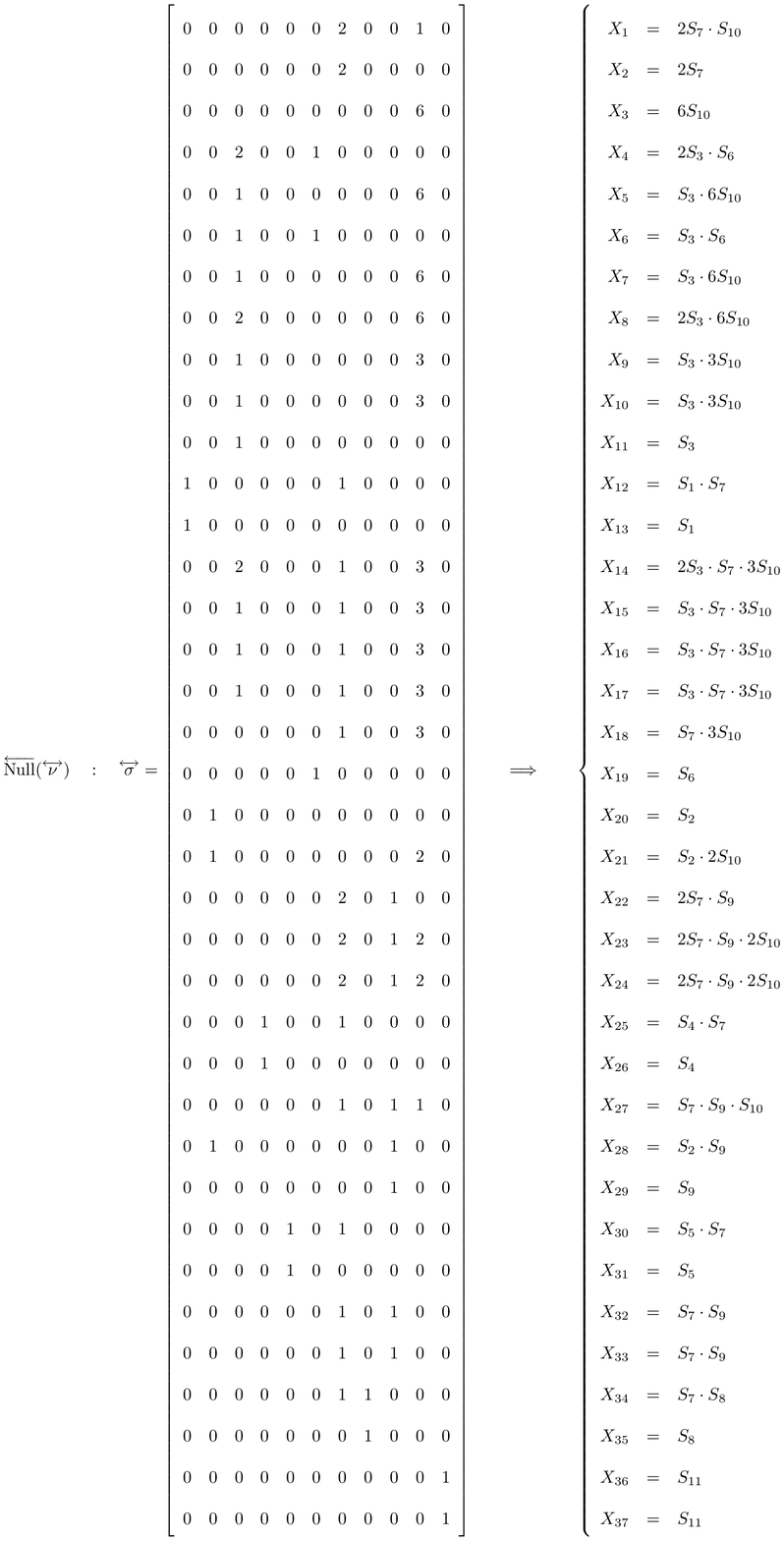}
      \caption{}
      \label{fig:ecoli-reacton}
   \end{center}
\end{figure}

A base of the left null space can be computed from this matrix,
leading to the molecular decomposition given in
Fig.~\ref{fig:ecoli-reacton}. The molecules are reduced to the
combination of the following reactons:
\begin{eqnarray}
  S_1 &=& NADH \\
  S_2 &=& COA \\
  S_3 &=& P_i \\
  S_4 &=& NADPH \\
  S_5 &=& FADH \\
  S_6 &=& AMP \\
  S_7 &=& O \\
  S_8 &=& QH_2 \\
  S_9 &=& SUCC \\
  S_{10} &=& C \\
  S_{11} &=& H
\end{eqnarray}
We must note that, in the equations, the water molecules are
implicit, and thus do not appear in this decomposition. This
explains why the glucose ($X_3$) is written as $6S_{10}=C_6$ rather
than $\mathrm{C_6H_{12}O_6}=\mathrm{C_6(H_2O)_6}$.

We can see here how the $37$ chemical compounds of the network can
be reduced to a combination of only $11$ reactons. If this
decomposition is an obvious one for any biochemist, it is
important to understand here that it can be automatically
obtained, with no further knowledge that the stoichiometric
matrix. On the basis of this reacton decomposition, it is then be
possible to focus on the evolution of some given reacton, e.g. the
evolution of $C$ in the metabolism, from glucose to carbon
dioxide, the evolution of $O$ from dioxygen to carbon dioxyde, the
use of $P_i$ throughout the whole network, etc.

\begin{figure}
   \begin{center}
      \includegraphics*[width=6.5in]{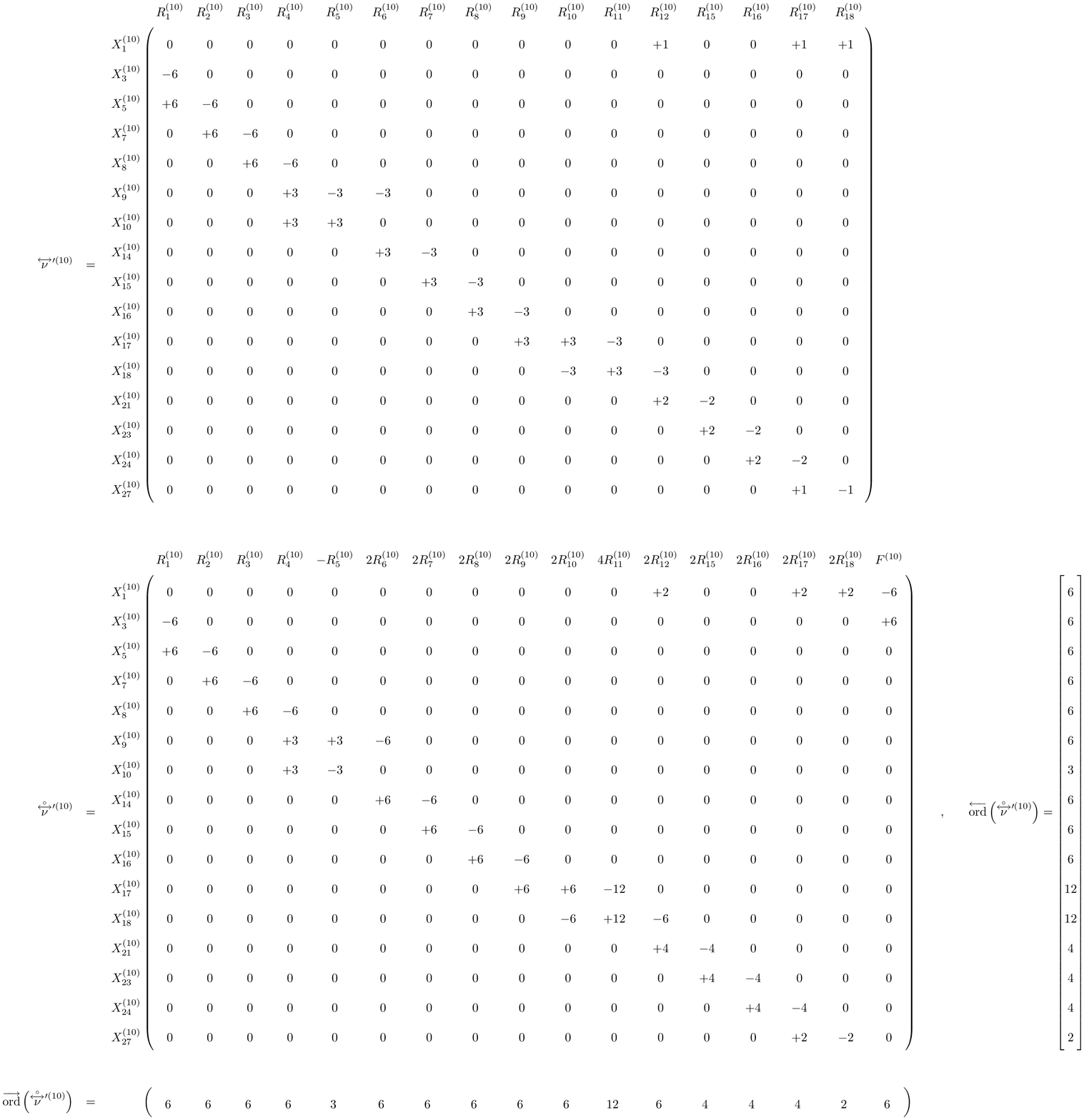}
      \caption{}
      \label{fig:ecoli-reacton10}
   \end{center}
\end{figure}

For example, following the reacton $S_{10}$, that is the carbon
coming from glucose, we can reduce the whole stoichiometric matrix
$\overleftrightarrow{\nu}$ of dimension $37 \times 28$ to the
stoichiometric matrix of the sub-network relative to $S_{10}$
$\overleftrightarrow{\nu}'^{(10)}$ of dimension $16 \times 16$
(see Fig.~\ref{fig:ecoli-reacton10}). It can then easily be
decomposed into single fluxes and cycles (see
Fig.~\ref{fig:ecoli-c-fluxes}). It becomes more tractable to
identify the progressive degradation of glucose, each carbon
following a linear flux towards carbon dioxide and being released
in three possible places. The whole system is coupled to the
PEP/PYR cycles.

\section*{Conclusion}

We have shown in this article that the left null space analysis of
the stoichiometric matrix can lead to an automatic decomposition
of molecules into physically meaningful sub-elements called
"reactons". Besides giving insight to the different moieties that
can be studied through the network, the discovery of reactons
leads to a natural simplification of the network, by dividing it
into subnetworks, each one related with one specific reacton.
These subnetworks can be easily studied and understood in
terms of simple fluxes and loops, by separating the reactions into
unitary partial reactions, describing the transfer of one reacton
from one molecule to another. The global network can then be seen
as a coupling of these elementary sub-elements. All these
algorithmic manipulations can be grouped into one single software
that remains simple to implement and use. Once the reactons have
been identified and the corresponding decomposition into
subnetworks achieved, the different modes are readily obtained.

However, the simplification allowed by such a decomposition into
reactons is nevertheless offset by the difficulty of deriving
an optimal reacton decomposition. This amounts to
computing a sparse null space base, which is far from being a
trivial problem~\citep{coleman-86,coleman-87}. This step remains
however of fundamental importance, as the sparsest null space will
lead to the largest interesting reactons and the simplest
corresponding sub-networks. The ``brute-force'' approach -- i.e.
computing a null space from the classical Gauss-Jordan
elimination~\citep{press-92} -- is easy to implement, but only
leads to an optimal solution following a huge amount of possible
linear combinations of vectors. This approach is not realistic for
large systems on account of the exponentially increasing number of
operations. This problem has nevertheless been thoroughly studied in
the literature~\citep{coleman-86,coleman-87,berry-85,gilbert-87},
and a careful examination of such work should help the future
development of new algorithms that are better adapted for obtaining
the optimal reactons.




\bibliography{biblio}

\clearpage
\section*{Figure Legends}

\subsubsection*{Figure~\ref{fig:global-loop}.}
Representation of the example chemical network given in
Eq.~\ref{eq:ex1}-\ref{eq:ex3}.

\subsubsection*{Figure~\ref{fig:mat-decomp}}
Decomposition of
$\accentset{\circ}{\overleftrightarrow{\nu}}^{(3)}$. a) Detail of the
operations leading to a square matrix representing unitary
transfers of reactons from one molecule to another. b) Detail of the
operations leading to the decoupling of cycles into elementary
independent cycles.

\subsubsection*{Figure~\ref{fig:loops}}
Graphical representation of the flux/cycle decomposition of the
example chemical network of Fig.~\ref{fig:global-loop}. a) Complete
decomposition of the flux and cycle relative to the reacton $S_3$. The
partial reactions are represented in red, and the reactons are
represented in violet. b) Simplified representation of the flux and
cycle relative to the reacton $S_3$. The dots represent the coupling
between partial reactions inside the complete reactions. c) Flux/cycle
decomposition representation for the whole network. The flux of $S_1$
is in red, the flux of $S_2$ in green and the flux and cycles of $S_3$
in blue. The dots represent the link between partial reactions, and
the segments represent the link between reactons.

\subsubsection*{Figure~\ref{fig:ecoli-metab}}
Chemical network representing a partial metabolism of
E.Coli~\citep{beard-04}, composed of 37 molecules and 28 reactions. 

\subsubsection*{Figure~\ref{fig:ecoli-stoich}}
Stoichiometric matrix of the E.Coli chemical network described in
Fig.~\ref{fig:ecoli-metab}. 

\subsubsection*{Figure~\ref{fig:ecoli-reacton}}
Sparse base of the left null space of the stoichiometric matrix of the
E.Coli chemical network of Fig.~\ref{fig:ecoli-stoich}.

\subsubsection*{Figure~\ref{fig:ecoli-reacton10}}
Matrix representations of the subnetworks of the E.Coli chemical
network, relative to the flow of organic carbon (reacton $S_{10}$). 

\begin{figure}
   \begin{center}
      \includegraphics*[width=6.5in]{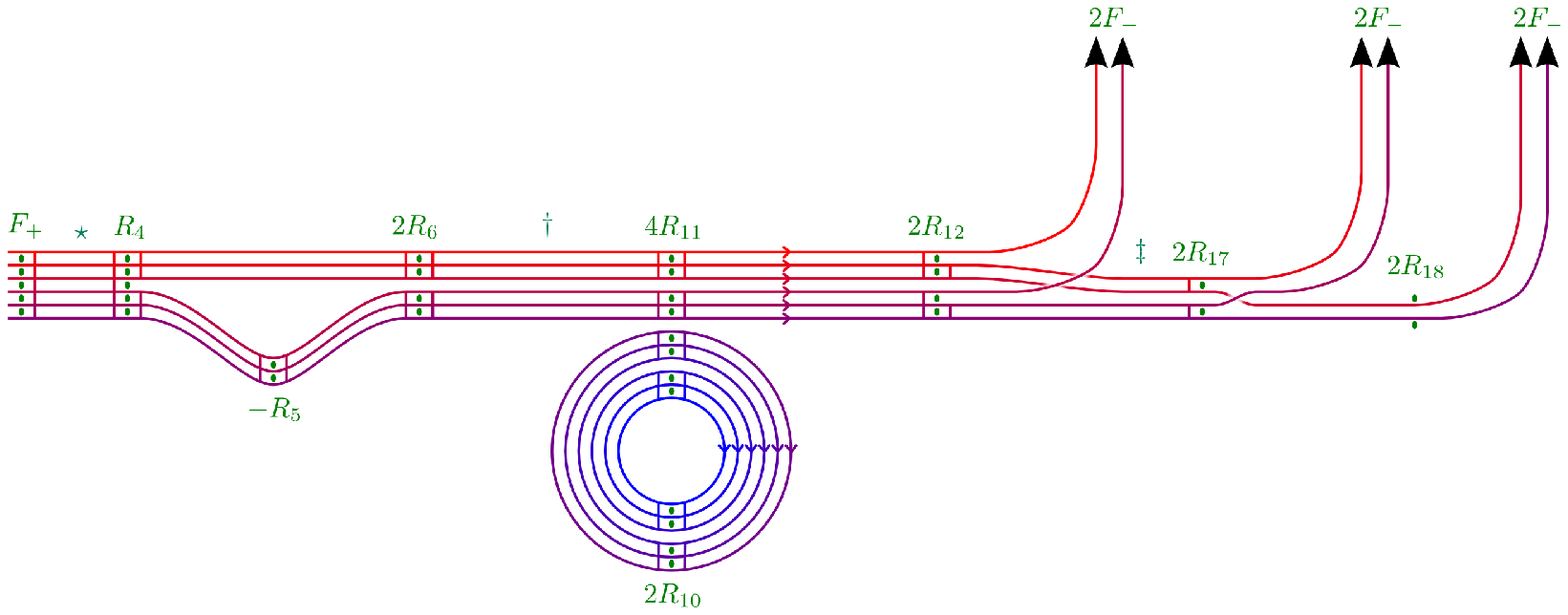}
      \caption{}
      \label{fig:ecoli-c-fluxes}
   \end{center}
\end{figure}

\subsubsection*{Figure~\ref{fig:ecoli-c-fluxes}}
Flux/cycle decomposition of the subnetworks of the E.Coli chemical
network, relative to the flow of organic carbon (reacton $S_{10}$).
The dots represent the link between partial reactions, and the
segments represent the link between reactons. $\star$: Reactions
$R_1$, $R_2$ and $R_3$.   $\dagger$: Reactions
$2R_7$, $2R_8$ and $2R_9$.  $\ddagger$: Reactions
$2R_{15}$ and $2R_{16}$.

\clearpage

\clearpage

\clearpage

\clearpage

\clearpage

\clearpage

\clearpage

\clearpage

\end{document}